\documentstyle[11pt,newpasp,twoside,epsfig]{article}
\def\edcomment#1{\iffalse\marginpar{\raggedright\sl#1\/}\else\relax\fi}
\reversemarginpar

\begin{document}
\title{A stellar fly-by simulation giving $\beta$ Pic's disc asymmetries}
\author{John Larwood}
\affil{Queen Mary \& Westfield College, Mile End Road, London E1 4NS, UK}
\author{Paul Kalas}
\affil{University of California, 601 Campbell Hall, Berkeley, CA 94709, USA}

\begin{abstract}
We have numerically investigated the dynamics of how a close stellar
fly-by encounter of a symmetrical circumstellar planetesimal disc
can give rise to the many kinds of asymmetries and substructures
attributed to the dusty disc of $\beta$ Pic. We find three recognizable
groupings of test particles that can be related to the morphology of
the $\beta$ Pic disc.
These are: highly eccentric and inclined orbit particles that
reach apocentre in the southwest, moderately eccentric and
inclined orbit particles that reach apocentre in the northeast,
and a relatively unperturbed region inside $\sim$200\,AU radius.
\end{abstract}

\section{Introduction}

The starlight of $\beta$ Pic scatters off dust grains in its
edge-on circumstellar disc (Smith \& Terrile 1984).
Imaging reveals that the disc exhibits asymmetrical structure
with respect to an ideal disc on projected radial scales from
$\sim$50\,AU (Burrows et al. 1995; Mouillet et al. 1997)
to $\sim$1000\,AU (Smith \& Terrile 1987). 
Sensitive $R$-band images (Larwood \& Kalas 2000) show the
northeast (NE) extension can be detected
out to $1835$\,AU, whereas the southwest (SW) extension is
detectable out to just $1450$\,AU, corresponding to a
$\sim$25 per cent length asymmetry.
More recently, the longer NE midplane extension has been discovered
to carry several brightness enhancements from
$\sim$500 to $800$\,AU (Kalas et al. 2000),
that have been attributed to a series of rings within the disc's
midplane. Similar features in the SW extension were not detected
and so eccentric rings were hypothesised.
Kalas et al. (2000) presented a dynamical model in which the stellar fly-by
scenario (Kalas \& Jewitt 1995) for
generating the disc asymmetries was applied to the $\beta$ Pic system.
Their finding was that the asymmetry types present in the $\beta$ Pic
disc have analogues in the dynamical response of a particle disc
to a perturbing stellar fly-by encounter. Furthermore, they
demonstrated that these features could occur simultaneously with
the formation of eccentric circumstellar rings.

As in the other works mentioned above
we assume that the simulation particles' distribution represents
the distribution
of an underlying disc of planetesimals, which are parent bodies
whose infrequent and therefore dynamically inconsequential collisions
supply the dust scattering surface that is observed in the
real system. A more complete model should include consideration
of grain removal and generation processes in determining the
appearence of the perturbed disc.
However, we defer treatment of those issues to future work,
and focus here on the first-stage problem of the dynamics.
This stage is important in deducing the mass and orbital parameters
of the postulated stellar perturber, for which it is possible to perform
star catalogue searches (Kalas, Deltorn \& Larwood 2000),
and we proceed to do this by
examining the length asymmetry, and other measures of the disc
response, as a function of stellar encounter parameters.

\section{Results}

Larwood \& Kalas (2000)
have described the origin of the transient circumstellar eccentric ring
structures as being a general outcome of an encounter, due to the reflex motion
of the primary star as the perturber passes through closest approach.
The apparent ring system is actually an eccentric tightly-wound one-armed
spiral pattern that gradually disappears owing to phase-mixing in the
particles' orbital motion. Here we summarise time-invariant
results found for the
encounter parameters favoured by Kalas et al. (2000). Namely, a
stellar perturber of mass 0.5\,M$_\odot$, moving in a prograde parabolic
trajectory with inclination of pericentre of 30$^\circ$. Fig. 1 shows
particle plots for some of the perturbed orbital elements in a disc
of $10^4$ test particles that were set up in Keplerian circular coplanar
orbits. The initial disc was set up between radii of 0.2 and 2, and
the distance of pericentre was 2.6. Kalas et al. (2000) suggested
a scaling such that the unit of distance corresponds to 270\,AU.

The disc response is separable into two major components corresponding
to the direct tidal interaction between disc particles and the
perturber, and the indirect interaction that results from the
reflex motion of $\beta$ Pic (Larwood \& Kalas 2000). This is evident
in Fig. 1 where we see, in the apocentre distance against longitude
of pericentre plot, oppositely
directed extensions that are asymmetric in length. The more diffuse
extension is recognisable in the other panels of Fig. 1, indicating
that these particles have high eccentricities and inclinations, and
small pericentre distances. This extension is formed by the direct
interaction. The other extension, formed by the indirect interaction,
has much lower eccentricity and inclination, and large pericentre
distances. Respectively, we identify these extensions with the SW
and NE extensions in the $\beta$ Pic disc.

\begin{figure}
\centerline{\epsfig{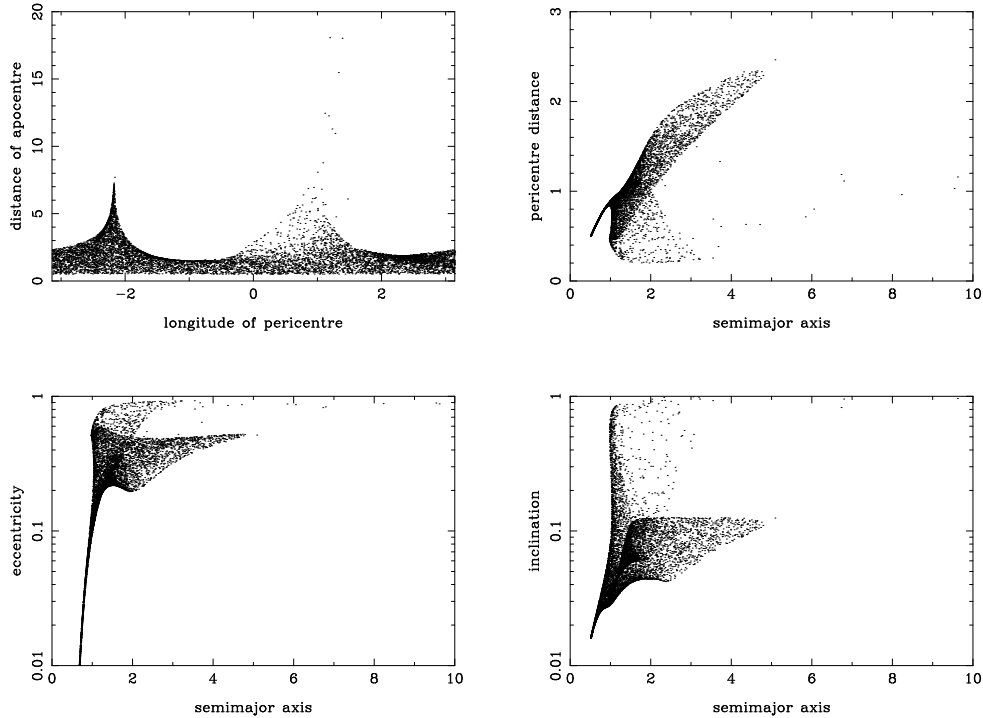}}
\caption{Orbital elements of perturbed disc particles.}
\end{figure}

\section{Conclusions}

The dynamical response of the models suggests that the $\beta$ Pic
disc might be separable into different components corresponding
to groupings of perturbed particle orbits. For example, we find that outwards
of $\sim$250\,AU the SW extension may not have a distinct midplane
owing to pumping of particle inclinations. The positions close to
pericentre of the moderately eccentric NE midplane particles could account
for the measurement of a SW midplane out to $\sim$650\,AU, in
superposition with the more diffuse vertically and radially extended component.
When the optical image is vertically compressed (Larwood \& Kalas 2000)
we are able to measure the
SW extension to much larger radii ($1450$\,AU). In our model the
corresponding (very eccentric
and inclined) particle orbits also intersect the disc near to their
pericentres (down to $\sim$50\,AU) possibly accounting for the inferred
inner warp of the disc, which is aligned with the outer flared
envelope in the SW. In summary, there are three recognizable particle
groupings that can be related to the morphology of the $\beta$ Pic disc.
These are: the highly eccentric and inclined particles that
reach apocentre in the SW, the moderately eccentric and low
inclination orbiting particles that reach apocentre in the NE,
and the relatively unperturbed particles inside $\sim$200\,AU radius.

\end{document}